\documentclass[12pt]{article}
\usepackage{amsmath}
\usepackage{graphicx}
\usepackage{amsfonts}
\usepackage{amssymb}
\begin{document}

\title{The anomalous Hall effect in ferromagnetic Fe: Skew scattering or side jump?}
\author{Gerd Bergmann and Manjiang Zhang\\Physics Department, University of Southern California,\\Los Angeles, CA 90089-0484, USA}
\date{\today}
\maketitle
\begin{abstract}
The question is investigated whether the anomalous Hall effect (AHE) in Fe
films is due to skew scattering or side jump. For this purpose sandwiches of
FeIn are investigated in which the conduction electrons carry their drift
velocity across the interface. This yields an additional AHE conductance
$\Delta G_{xy}$ whose dependence on the In mean free path is used to determine
the mechanism of the AHE in the Fe film. The structure of the Fe film is kept constant.

\end{abstract}

In ferromagnetic metals and metals with magnetic impurities one observes two
contributions to the Hall effect, (i) the normal Hall effect and (ii) the
anomalous Hall effect (AHE). The AHE is caused by spin-orbit scattering
through the interaction of the conduction electron spin with the magnetic
moments of the sample. The anomalous Hall effect was already observed by Hall
a century ago \cite{H28}. However, theoretically it is a rather complicated
problem. There are two main mechanisms discussed in the literature, (a) skew
scattering and (b) side jump. Both require a scattering mechanism for the
conduction electrons and vanish in pure samples. The first models of skew
scattering were developed by Karpulus and Luttinger \cite{L25} and Smit
\cite{S30}, while the side jump was proposed by Berger \cite{B138}. Due to its
importance in spintronics, the anomalous Hall effect has been intensively
studied in recent years \cite{N10}, \cite{S45}, \cite{K48}, \cite{O27} (for
further references see \cite{C19}). Recently an additional mechanism has been
under discussion which is connected with the Berry phase and believed to occur
even in the absence of any scattering (see for example \cite{K49}). Here we
restrict the discussion to the skew scattering and the side jump.

It is often stated in theoretical papers that for skew scattering the
anomalous Hall resistivity $\rho_{yx}$ is proportional to the resistivity
$\rho_{xx}$ while for the side jump $\rho_{yx}$ is proportional to the square
of the resistivity $\rho_{xx}^{2}$. A number of experimental investigation
tried to identify the mechanism of the AHE by changing the resistivity of
their sample and analyzing the dependence of $\rho_{yx}$ on $\rho_{xx}$.

In this paper we will first show that the power law $\rho_{yx}\varpropto
\rho_{xx}^{p}$ with $p=1$ for skew scattering and $p=2$ for the side jump is
rather poorly justified. In the second part of the paper we use a very
different experimental approach to identify the origin of the AHE in thin
amorphous Fe films.

We briefly review the AHE resistivity for skew scattering and the side-sump in
a ferromagnetic sample where the magnetic moments are aligned in the
z-direction. The smple is disordered and has a finite $\rho_{xx}$. Part of the
scattering will be potential scattering, i.e. spin-independent, and another
part will be magnetic or spin-dependent.

We begin with \textbf{skew scattering}. First we consider conduction electrons
with spin $\sigma$. In Fig.1 an electron propagates in the x-direction. A part
of the wave is skew-scattered by a magnetic moment. We describe the potential
scatterers by their concentration $n_{i}$ and their total scattering cross
section $a_{i}$ (The index $i$ for impurity). Similarly the magnetic
scatterers have the concentration $n_{m}$ with the scattering cross section
$a_{m\sigma}$, i.e. the strength of the scattered wave is given by
$a_{m\sigma}$ ($a_{m\sigma}$ depends on the spin of the conduction electron).
The integrated momentum of the skew scattered wave possesses an electron
momentum in the y-direction with the weight $a_{AH,\sigma}.$Here
$a_{AH,\sigma}$ is the AHE cross section. It is defined so that $\hbar
k_{F}a_{AH,\sigma}$ is equal to the y-component of the integrated momentum of
the scattered wave. (A possible forward scattering should be incorporated into
the scattering cross section $a_{m\sigma}$.) With the definition of a
relaxation time $\tau_{\sigma}$ and the corresponding mean free path
$l_{\sigma}$%
\[
\frac{1}{\tau_{\sigma}}=\frac{v_{F,\sigma}}{l_{\sigma}}=v_{F,\sigma}\left(
n_{i}a_{i}+n_{m}a_{m\sigma}\right)
\]
one obtains for the longitudinal and transverse resistivities $\rho
_{xx,\sigma}$ and $\rho_{yx,\sigma}$ for each spin direction $\sigma$%
\begin{align}
\rho_{xx,\sigma} &  =\frac{m}{n_{\sigma}e^{2}\tau_{\sigma}}\label{royx_skew}\\
\rho_{yx,\sigma} &  =\rho_{xx,\sigma}l_{\sigma}n_{m}a_{AH,\sigma}%
=\rho_{xx,\sigma}\frac{n_{m}a_{AH,\sigma}}{n_{i}a_{i}+n_{m}a_{m\sigma}%
}\nonumber
\end{align}

The ratio $\rho_{yx,\sigma}/\rho_{xx,\sigma}$ (and therefore $\rho_{yx,\sigma
}$) is rather intuitive. According to Fig.1 the electrons propagate the
distance of the mean free path $l_{\sigma}$ (MFP) before they lose their drift
velocity. After the skew scattering the fraction $n_{m}a_{AH,\sigma}/\left(
n_{i}a_{i}+n_{m}a_{m\sigma}\right)  $ of the electron propagates the same MFP
$l_{\sigma}$ in y-direction. This yields the ratio $l_{y,\sigma}/l_{x,\sigma
}=\rho_{yx,\sigma}/\rho_{xx,\sigma}=n_{m}a_{AH,\sigma}/\left(  n_{i}%
a_{i}+n_{m}a_{m\sigma}\right)  $.%
\[%
\begin{tabular}
[c]{l}%
{\includegraphics[
height=3.8032in,
width=3.3059in
]%
{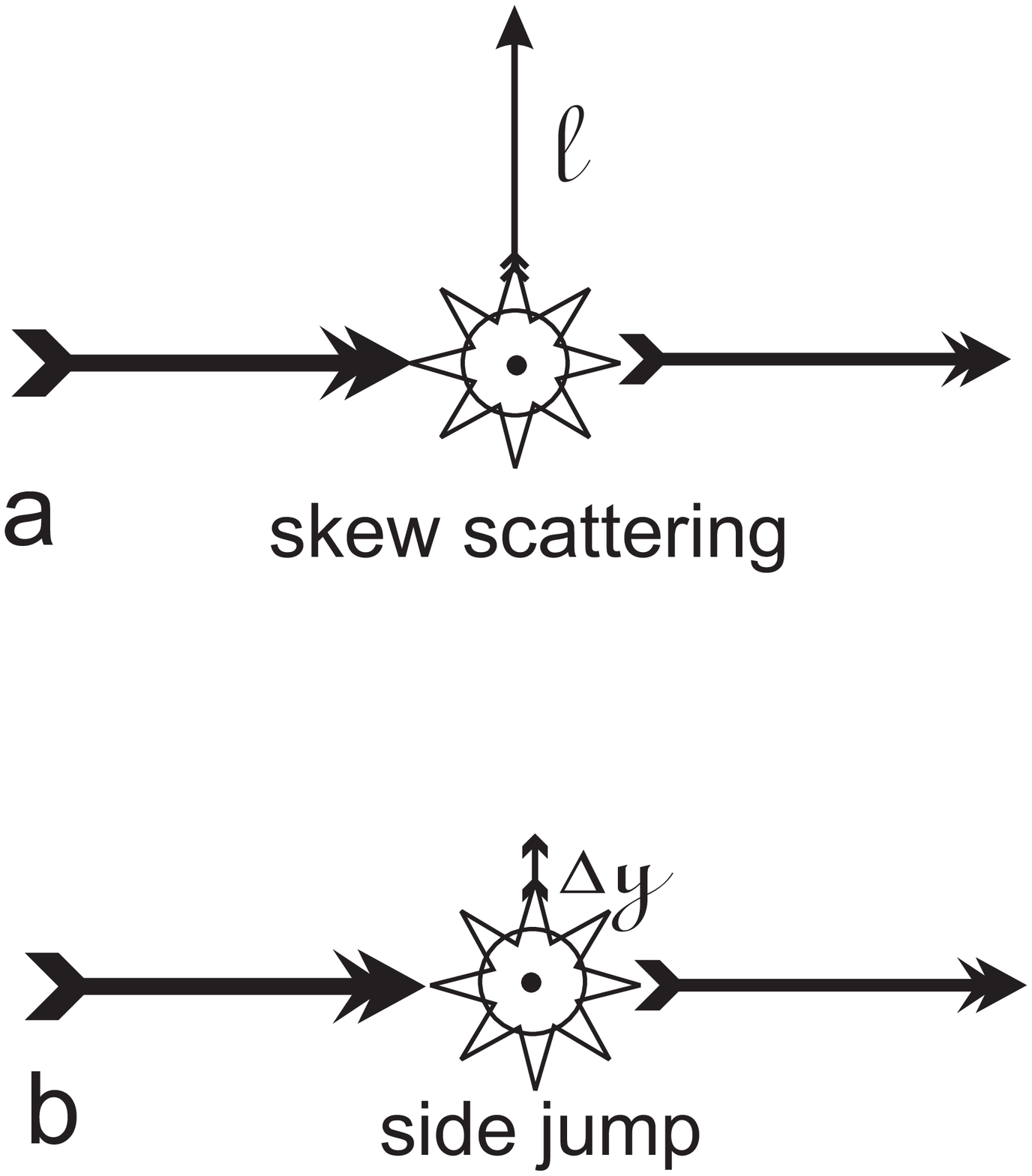}%
}%
\\
Fig.1a: An (spin up) electron wave with momentum $\mathbf{k=}\left(
k_{x},0,0\right)  $\\
propagates in the x-direction. A part $a_{m\sigma}$ of the wave is
skew-scattered\\
by a magnetic moment and carries a momentum in the y-direction.\\
b) This time the side jump displaces the scattered electrons wave\\
by $\Delta y_{\sigma}$ in the y-direction.
\end{tabular}
\ \ \ \
\]

Equation (\ref{royx_skew}) yields the well known statement that the AHE
resistivity $\rho_{yx,\sigma}$ is proportional to the resistivity
$\rho_{xx,\sigma}$, \textbf{for a single spin direction! }Since the electron
has two spins we have to (i) invert the resistivity tensors $\left(  \rho
_{ij}\right)  _{\uparrow,\downarrow}$ for each spin to obtain the conductivity
tensors $\left(  \sigma_{ij}\right)  _{\uparrow,\downarrow}$, (ii) add the two
conductivity tensors and (iii) invert the resulting tensor.
\begin{equation}
\left(  \rho\right)  =\left(  \left(  \rho\right)  _{\uparrow}^{-1}+\left(
\rho\right)  _{\downarrow}^{-1}\right)  ^{-1} \label{invro}%
\end{equation}
Since the MFPs of spin up and down electrons in the ferromagnet are generally
quite different the original linearity between $\rho_{yx,\sigma}$ and
$\rho_{xx,\sigma}$ for the individual spin is replaced by a complicated dependence.

For the \textbf{side jump} the electron does not propagate in the y-direction
after the scattering but the whole scattered electron is displaced by the
distance $\Delta y_{\sigma}$ in the y-direction. For the event sketched in
Fig.1 the propagation in the x-direction is again $l_{x,\sigma}=l_{\sigma}$
while the "propagation" in the y-direction is equal to $l_{y,\sigma}=$ $\Delta
y_{\sigma}n_{m}a_{m\sigma}/\left(  n_{i}a_{i}+n_{m}a_{m\sigma}\right)  $. This
yields for $\rho_{yx,\sigma}$ (for each spin) the following value for the side
jump
\begin{equation}
\rho_{yx,\sigma}=\rho_{xx,\sigma}\frac{l_{y,\sigma}}{l_{\sigma}}%
=\rho_{xx,\sigma}\Delta y_{\sigma}n_{m}a_{m\sigma} \label{royx_side}%
\end{equation}
If $\rho_{xx,\sigma}$ is proportional to density of magnetic scattering
centers $n_{m}$ then $\rho_{yx,\sigma}$ is proportional to the square of
$\rho_{xx,\sigma}$ if the scattering cross section $a_{m\sigma}$ and the side
jump are independent of the resistivity, \textbf{for a single spin direction!}
Using equ. (\ref{invro}) for the total AHE resistivity destroys the quadratic
relation. Furthermore the parameters $a_{i}$, $a_{m\sigma},a_{AH,\sigma}$
$\Delta y_{\sigma}$ are not independent of the disorder. The scattering
potential is generally not the atomic potential but the deviation from the
periodic potential. This potential is generally not spherically symmetric but
is rather the gradient of a spherical potential. All of the scattering
parameters $a_{i}$, $a_{m\sigma}$, $a_{AH,\sigma}$ and $\Delta y_{\sigma}$
will change in complicated ways with the disorder or alloying of the
ferromagnet. In particular the behavior of \ $\Delta y_{\sigma}$ as a function
of disorder is very critical. As Berger showed the side jump, which is caused
by the spin-orbit interaction, only becomes significant because the spin-orbit
interaction in the magnetic atoms can be enhanced by a factor of
$3\times10^{4}$. Even the smallest change in the local environment of the
magnetic atom could change this enhancement factor.

We summarize: The simple power law dependence of $\rho_{yx}$ on the
resistivity $\rho_{xx}$ for skew scattering and side jump may not be reliable
for the following reasons:

\begin{itemize}
\item  The contribution of two kinds of electrical carriers in ferromagnets,
spin up and down electrons, destroys the simple relation between $\rho_{yx}$
and $\rho_{xx},$

\item  The scattering parameters $a_{i}$, $a_{m\sigma}$, $a_{AH,\sigma}$ and
$\Delta y_{\sigma}$ will change in complicated ways with the disorder or
alloying of the ferromagnet.

\item  The enhancement of the spin-orbit interaction (which determines the
side-jump parameter) by a factor of $3\times10^{4}$ will be very sensitive to
the disorder.
\end{itemize}

In the present investigation we use a new approach to investigate the AHE of a
ferromagnetic film. We use a thin ferromagnetic film as the target of a
scattering experiment by exposing it to incident electrons. The momentum of
the incident electrons is varied and the electrons are scattered by the
target. Their (integrated) angular scattering is measured. This appears to be
a conventional scattering experiment but there is an important difference. The
probing electrons are the conduction electrons of a normal metal film which is
condensed on top of the ferromagnetic film. The sandwich is shown in Fig.2. In
the experiment we use amorphous Fe with a very short MFP for the ferromagnet
and In with a much larger MFP for the normal metal. This simplifies the
underlying physics and the evaluation of the experiment.
\[%
\begin{tabular}
[c]{l}%
{\includegraphics[
height=1.6364in,
width=3.7592in
]%
{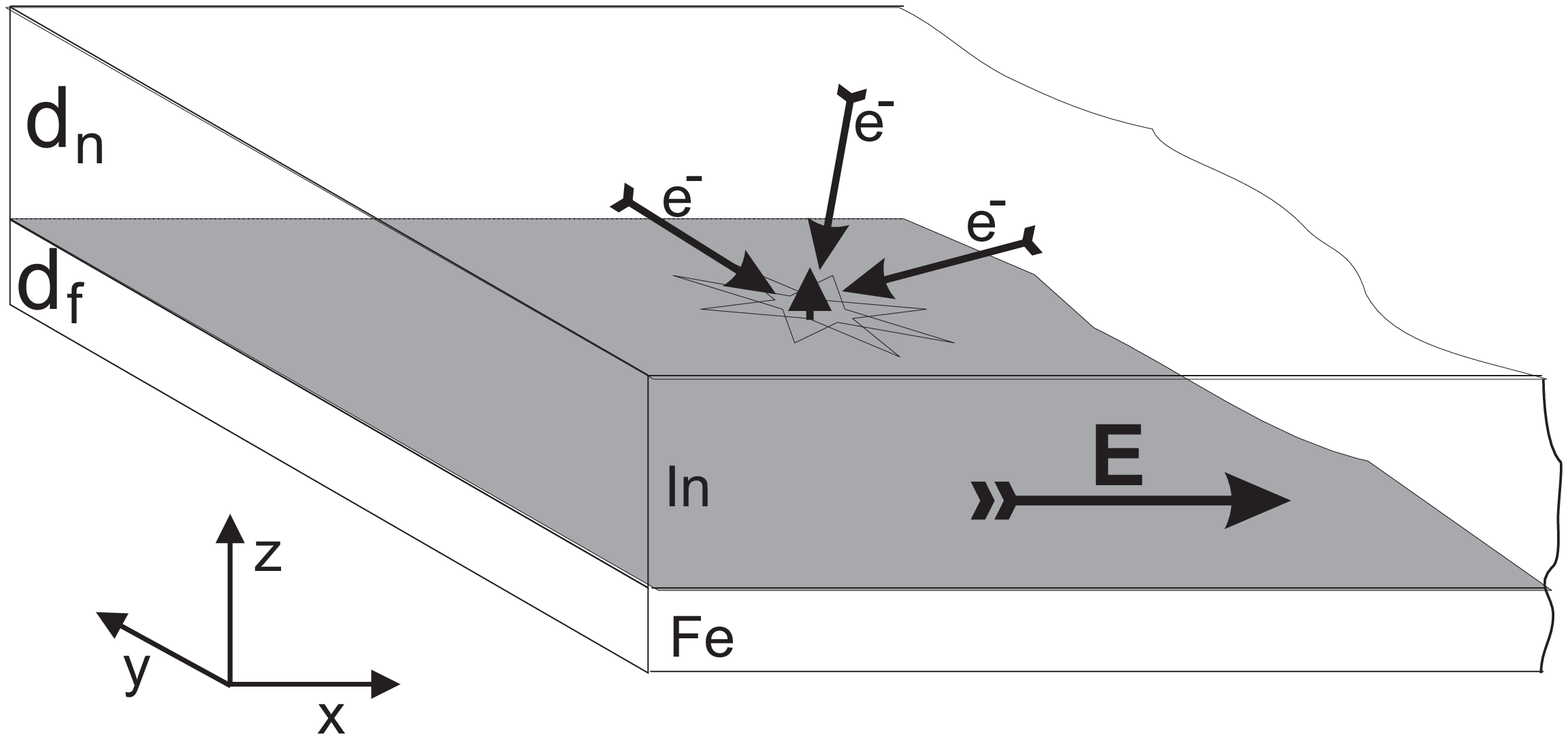}%
}%
\\
Fig.2: A sandwich consisting of ferromagnetic amorphous Fe and\\
the normal metal In is quench condensed. In the presence of an\\
electric field the conduction electrons in the In carry their larger\\
drift velocity into the upper layer of the Fe and create a large\\
anomalous Hall effect (AHE) in the Fe. Its dependence on the mean free\\
path of the conduction electrons in the In identifies the origin\\
of the AHE.
\end{tabular}
\ \ \ \ \
\]

In the presence of an electric field $E$ (in the x-direction) the electrons
accumulate finite drift velocities: $v_{n}$ in the normal metal and
$v_{f\uparrow},v_{f\downarrow}$ in the amorphous ferromagnet. The electrons of
both metals cross the interface. The electrons which cross from the normal
metal into the ferromagnet increase the current density in the upper layers of
the Fe dramatically because they carry a much larger drift velocity. This
injected high current density in the ferromagnet is proportional to the MFP in
the normal metal. It creates an additional AHE in the Fe. If the AHE is due to
the side-jump mechanism then the injected current yields an AHE conductance
which is proportional to the MFP $l_{n}$ \ in the normal metal. If the AHE is
due to skew scattering then a large fraction of the scattered electrons
returns into the normal metal and propagates there the distance $l_{n}$.
Therefore their contribution to the AHE conductance is proportional to the
square of the MFP in the normal metal. By changing the MFP in the normal metal
we can analyse the origin of the AHE in the ferromagnet without changing the
structure of the ferro-magnet.

Our FeIn sandwiches are prepared at liquid helium temperatures. To obtain very
flat and homogeneous Fe films we first condense 10 atomic layers of insulating
amorphous Sb. On this fresh substrate the Fe film shows conductance already
for one mono layer. The thickness of the Fe films lies in the range of 5 to 10
atomic layers. On top of the Fe film the In is condensed in several steps up
to a thickness of $25nm$. The MFP of the In lies in the range of $5-20nm$
while the MFP of the Fe is of the order of a few Angstroms. Fig.3 shows the
anomalous Hall curves for a sandwich of 5 atomic layers of amorphous Fe
covered with increasing layers of In. The normal Hall conductance is
subtracted. From these curves we obtain the AHE conductance $L_{xy}^{AHE}$ by
back extrapolation of the high field part of the curve to zero magnetic field.
We observe an increase of the AHE conductance with the In thickness. We denote
the additional AHE conductance as \emph{interface} AHE conductance.%

\begin{align*}
&
\raisebox{-0pt}{\includegraphics[
height=2.9805in,
width=4.3462in
]%
{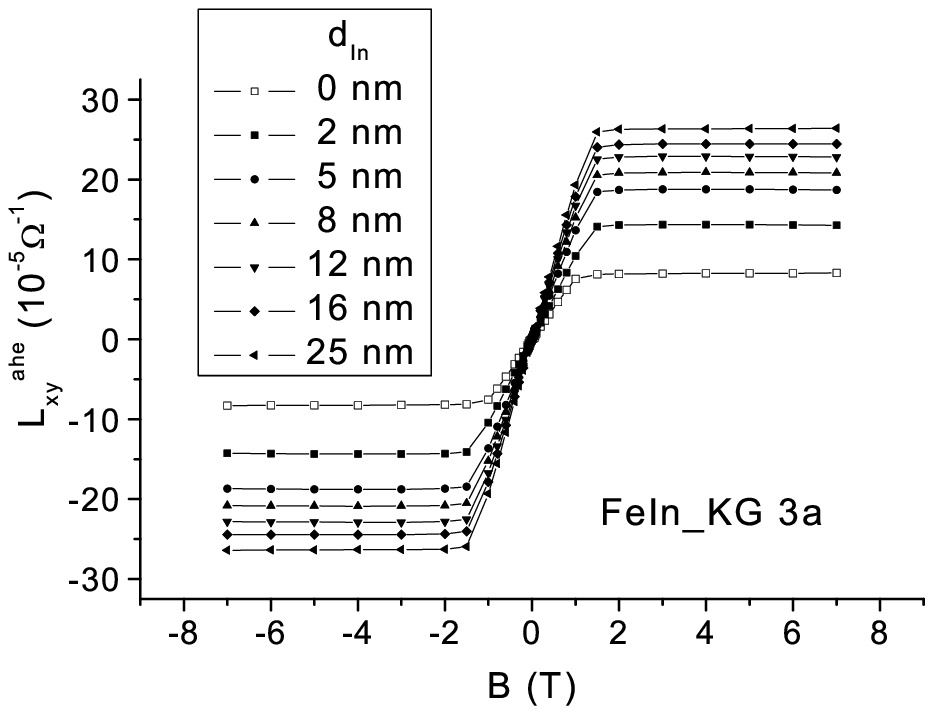}%
}%
\\
&
\begin{tabular}
[c]{l}%
Fig.3: The anomlous Hall conductance curves for a thin Fe film\\
covered with In of increasing thickness.
\end{tabular}
\end{align*}

In Fig.4 we have plotted the interface AHE conductance $\Delta L_{xy}$ as a
function of the In thickness for two different amorphous Fe film thicknesses
of 5 and 8.6 atomic layers. The fact that the two curves lie very close to
each other demonstrates that the interface AHE conductance does not depend on
the thickness of the ferromagnet (as long as the thickness is larger than the
MFP).
\begin{align*}
&
{\includegraphics[
height=3.2594in,
width=4.0598in
]%
{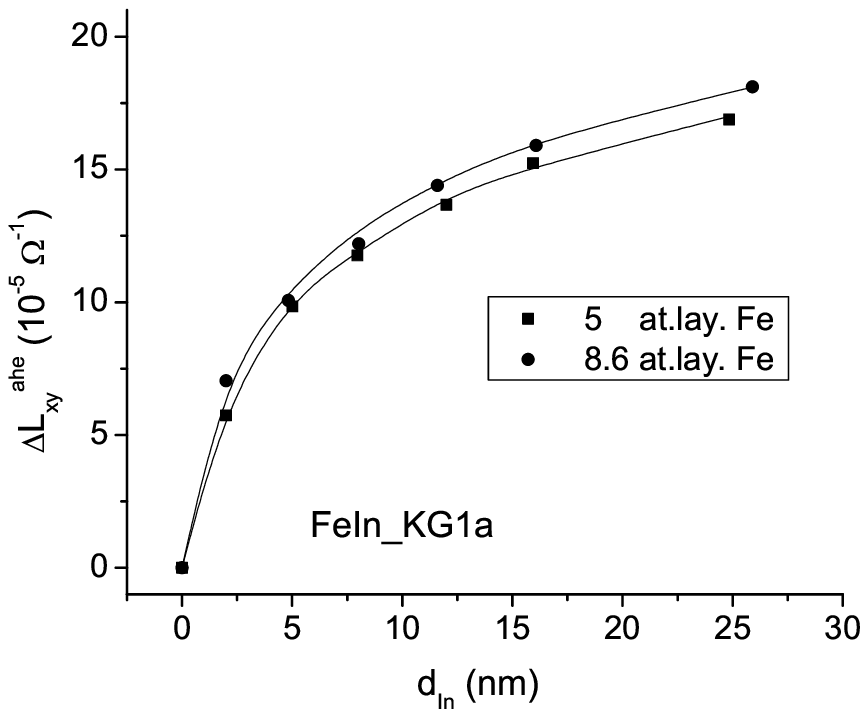}%
}%
\\
&
\begin{tabular}
[c]{l}%
Fig.4: The interface anomalous Hall conductance of two FeIn\\
sandwiches as a function of the In thickness. The thicknesses\\
of the amorphous Fe films are 5 and 8.6 atomic layers.
\end{tabular}
\end{align*}

For the evaluation and interpretation of the experiment we calculated the AHE
conductance of an FN (ferromagnet/non-magnetic metal) sandwich. (The details
will be published elsewhere). We applied the (linearized) Boltzmann equation
using Chamber's method of the vector mean free path \cite{C17}. The fact that
the Fermi energies differ in the two metals complicates the calculation
considerably. Therefore we follow here the examples in the theory of giant
magneto-resistance and the superconducting proximity effect where the first
theoretical approaches simplified the problem by assuming identical electronic
properties in both metals. Furthermore we take the densities of spin up and
down electrons in the ferromagnet as identical and equal to $n/2$.

The thickness and MFPs of the Fe film are denoted as $d_{f},l_{f\uparrow},$
$l_{f\downarrow}$for spin up and down and $d_{n},l_{n}$ for the In. Because
the Fe and In film are in parallel their conductances would simply add if
there would be no interface crossing between the films. Without the crossing
the In would not contribute to the AHE.

The conduction electrons in the In with $k_{z}<0$ cross through the interface
into the Fe. In the following we call them the injected electrons. In the Fe
they carry an injected current $I_{in}$ for each spin which for $l_{n}%
>>l_{f\uparrow},l_{f\downarrow}$ is
\[
I_{in}=\frac{1}{16}e^{2}N_{0}v_{F}l_{n}\left(  l_{f\uparrow}+l_{f\downarrow
}\right)  E
\]
where $N_{0}$ is the density of electron states per spin. This current flows
in a thin layer of the Fe whose thickness is half the MFP, i.e.,
$l_{f\uparrow}/2$ or $l_{f\downarrow}/2$ for spin up and down electrons and is
proportional to the MFP in the normal metal. The result for the longitudinal
part of the conductance $G_{xx}=I_{in}/E$ is similar to Fuchs \cite{F31} and
Sondheimer \cite{S36} for thin films but extended to sandwiches. The injected
current yields an additional large AHE. The resulting contribution to the
anomalous Hall conductance depends on the mechanism of the AHE.

\textbf{side jump:} The electrons which cross from the normal metal into the
ferromagnet contribute to the side jump. They yield an additional AHE
conductance%
\[
\Delta G_{xy}^{\left(  sj\right)  }=\frac{1}{16}e^{2}N_{0}v_{F}l_{n}\left(
l_{f\uparrow}\Delta y_{\uparrow}n_{m}a_{0\uparrow,m}+l_{f\downarrow}\Delta
y_{\downarrow}n_{m}a_{0\downarrow,m}\right)
\]
The electrons which cross from the ferromagnet to the normal metal do not
contribute to the AHE.

\textbf{skew scattering:} In contrast to the side jump, here part of the
important physics happens after the scattering because half of the skew
scattered electrons propagate back towards the normal film. Therefore one
obtains two additional contributions due to the normal metal film: (i) The
conduction electrons which are accelerated in the ferromagnet and cross into
the normal metal after the scattering and (ii) the conduction electrons which
are accelerated in the normal metal, cross the interface into the ferromagnet,
experience skew scattering and then cross back into the normal metal. The
second effect is proportional to $l_{n}^{2}$ and is dominant. It yields an
additional anomalous Hall conductance
\[
\Delta G_{xy}^{(ss)}=\frac{\beta}{32}e^{2}N_{0}v_{F}l_{n}^{2}n_{m}\left(
a_{ah\uparrow}l_{f\uparrow}+a_{ah\downarrow}l_{f\downarrow}\right)
\]
The additional factor $\beta/2$ (with respect to the current) is due to the
fact that only half of the scattered electrons move back towards the normal
metal. Since they are roughly the distance $l_{f\uparrow}/2,l_{f\downarrow}/2$
from the interface only the fraction $\beta$ reaches the normal metal without
being scattered in the ferromagnet. The factor $\beta$ is less than one and of
the order of $1/2$. (If the scattering in the ferromagnet would be isotropic
then $\beta$ would have the value 1/2)

For skew scattering the additional anomalous Hall conductance is proportional
to the square of the MFP in the normal metal. This result is quite physical.
The interface AHE conductance is (i) proportional to the drift velocity, i.e.
$l_{n}$, and (ii) proportional to the distance the electrons travel after the
scattering, which yields another factor of $l_{n}$.

For the analysis of the experiment we plot in Fig.5 the total anomalous Hall
conductance of the two sandwiches versus the MFP of the conduction electrons
in the In film. Together with the experimental points is shown a linear fit.
Obviously the interface AHE conductance is linear in the MFP $l_{n}$ of the
In. This proves clearly that the anomalous Hall effect in the amorphous Fe
film is due to the side jump. Originally Berger's argument for observing the
side jump in metals with small MFP was that there the small MFP reduces the
magnitude of the skew scattering so that one could observe the side jump. Our
conclusion goes beyond that suggestion. Since the propagation in the normal
metal yields a contribution which is proportional to the square of the MFP in
the normal metal our experiment should detect a contribution of the skew
scattering even if its contribution in the amorphous Fe would be very small.
So the conclusion of our experiment is that there is practically no skew
scattering in the amorphous Fe.%
\begin{align*}
&
{\includegraphics[
height=3.1631in,
width=4.0598in
]%
{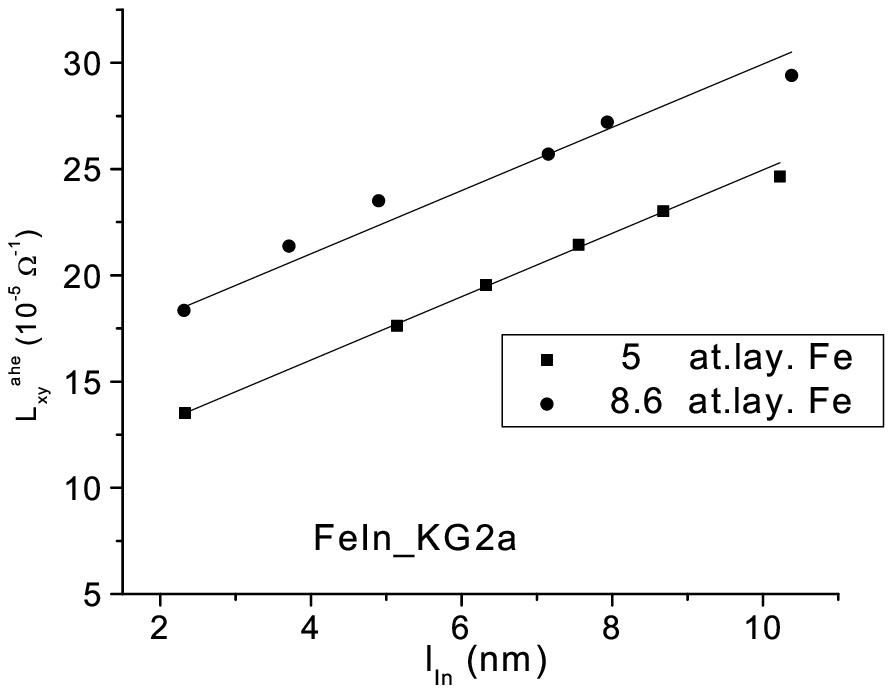}%
}%
\\
&
\begin{tabular}
[c]{l}%
Fig.5: The total anomalous Hall conductance of the two FeIn\\
sandwiches shown in Fig.5 as a function of the In mean free path.
\end{tabular}
\end{align*}

In this paper we have investigated the mechanism of the AHE in amorphous Fe.
The standard approach assumes that $\rho_{yx}$ depends on $\rho_{xx}$ linearly
for skew scattering and quadratically for side jump. First \ we pointed out
that this assumption is not well justified. Instead we introduced a method for
which the structure of the ferromagnet is kept constant. By preparing a
sandwich of amorphous Fe with the non-magnetic metal In we observed an
increased AHE conductance because conduction electrons with a larger drift
velocity cross from the In into the Fe and cause an additional "interface" AHE
within the MFP of the Fe. For the side jump this interface AHE conductance
$\Delta G_{xy}$ is proportional the MFP $l_{In}$ of the In. If the AHE is due
to skew scattering then about $1/4$ of the skew-scattered electrons cross back
into the In and propagate the distance $l_{In}$. This yields a quadratic
dependence of $\Delta G_{xy}$ on $l_{In}$ for skew scattering. The great
advantage of the interface AHE is that the structure and scattering potentials
of \ the ferromagnet are kept constant.

Our experimental results yield a linear dependence of $\Delta G_{xy}$ on the
In MFP $l_{In}$. This not only shows that the side jump is the dominant
mechanism for the AHE in amorphous Fe, but the experiment did not detect any
skew scattering in the amorphous Fe film.

Abbreviations: AHE=anomalous Hall effect, MFP=mean free path.

Acknowledgment: The research was supported by NSF Grant No. DMR-0124422.
\end{document}